\documentclass[aip,jcp,letterpaper,reprint,twocolumn,amsmath,amssymb,amsfonts,floatfix]{revtex4-1}

\usepackage{color}
\usepackage{times}
\usepackage{mathptmx}

\usepackage{graphicx}
\usepackage{braket}

\newcommand{\figref}[1]{Fig.\,\ref{#1}}

\begin{document}

\title{{\color{blue}Density matrix embedding: A strong-coupling quantum embedding theory}\vspace*{0.18cm}}
\author{Gerald Knizia}
\author{Garnet Kin-Lic Chan}
\affiliation{Department of Chemistry, Princeton University, Princeton NJ 08540}
\begin{abstract}
   We extend our density matrix embedding theory (DMET) [Phys.\;Rev.\;Lett. {\bf109}
   186404 (2012)] from lattice models to the full chemical Hamiltonian. DMET
   allows the many-body embedding of arbitrary fragments of a quantum system,
   even when such fragments are open systems and strongly coupled to their
   environment (e.g., by covalent bonds). In DMET, empirical approaches to
   strong coupling, such as link atoms or boundary regions, are replaced by a
   small, rigorous quantum bath designed to reproduce the entanglement between a
   fragment and its environment. We describe the theory and demonstrate its
   feasibility in strongly correlated hydrogen ring and grid models; these are
   not only beyond the scope of traditional embeddings, but even challenge
   conventional quantum chemistry methods themselves. We find that DMET
   correctly describes the notoriously difficult symmetric dissociation of a
   4$\times$3 hydrogen atom grid, even when the treated fragments are as small
   as single hydrogen atoms. We expect that DMET will open up new ways of
   treating of complex strongly coupled, strongly correlated systems in terms of
   their individual fragments.
\end{abstract}

\maketitle

  Embedding techniques are powerful tools for enabling high-level many-body
  treatments on system sizes they cannot normally reach. They work by dividing a
  chemical system into fragments, each of which is handled individually; the
  interaction with the other fragments---the environment---is treated in a
  simplified way. In this communication we are concerned with embeddings for
  fragments which are \emph{strongly} coupled to their environment, for example
  via covalent bonds.

  A particular embedding is characterized by the precise manner in which the
  environment, and its influence on the fragment, are represented. To date, most
  techniques represent the environment through a one-particle embedding
  potential $v$. For example, in QM/MM methods, $v$ is obtained through
  electrostatics or polarization
  interactions,\cite{Truhlar:QMMMreview,Thiel:QMMMreview} while in
  {\it ab-initio} DFT embedding, $v$ is the derivative of the non-additive
  energy functional.%
  \cite{Cortona:DftEmbedding91,Wesolowski:FrozenDenistyEmbedding,Huang:UniqueEmbeddingPotential,Goodpaster:ExactKineticPotentialsForEmbedding}
  However, an embedding potential, regardless of how it is obtained, cannot
  represent the effect of the environment on the many-body fragment state when
  the coupling is \emph{strong}.%
  \footnote{Note we are careful to write ``many-body'' fragment state. A
            potential {\it is} formally sufficient to represent the effect of
            the environment on the single-particle {\it density}, and this is
            the basis of exact DFT embedding. However, we assume here that we
            wish to describe the fragment with a high-level {\it many-body}
            treatment, not only at the level of the density. In this case, the
            environment cannot be represented by a potential.}
  This is illustrated by the simple example of embedding a hydrogen atom $A$
  within a hydrogen molecule $AB$. If hydrogen atom $B$ is represented by an
  embedding potential $v$, then hydrogen atom $A$ appears as a closed system
  with a single electron; thus, any wavefunction description of the fragment,
  regardless of the choice of $v$, provides no information on electron
  correlation.

  This failure of potential based embedding is rooted in the fact that the
  fragments are {\it open}  systems that are  entangled with their environment.
  Is it  possible {\it even in principle} to formulate an embedding description
  of an open fragment? This question has far reaching consequences; an
  affirmative answer would imply, for example, that one could in principle
  exactly calculate the properties of a bulk diamond crystal by treating a
  single embedded carbon atom. Here we argue that this is in fact the case. The
  key is to represent the environment not by a potential, but rather through a
  quantum {\it bath} that reproduces the entanglement of the embedded fragment
  with the full environment. In existing embedding approaches, empirical link
  atoms or boundary regions\cite{slavivcek:MCVEEP} can be thought of as baths,
  but here we show that an \emph{exact} bath, that exactly reproduces \emph{all}
  many-body environment effects, can in fact be defined. In realistic systems,
  the construction of this exact bath is not practical. However, as we show
  below, this point of view naturally leads to a practical embedding method
  which we call {\it density matrix embedding theory} (DMET).

  We have previously introduced DMET in the context of fermionic lattice
  models\cite{knizia:DmetHubbard} where it showed excellent performance; in
  particular, in comparison to the more complex dynamical mean-field theory
  (DMFT). Here we describe the extension and modification of DMET to treat
  inhomogeneous systems and the full chemical Hamiltonian with long-range
  interactions, namely
  \begin{align}\label{eq:QCH}
     H = \sum_{rs} h^r_s E^s_r + \frac{1}{2} \sum_{rstu} V^{rs}_{tu} E^{tu}_{rs}.
  \end{align}
  We note especially that DMET is defined through simple linear algebra and
  avoids the numerical issues of inverse problems associated with constructing
  potential embeddings, as found in \emph{ab-initio} DFT embedding
  (see also Ref. \onlinecite{manby:ProjectorEmbedding}, another solution to the
  inverse problem).

  We use the following notation: The full quantum system $Q$ is spanned by an
  orthogonal one-particle basis (e.g., an symmetrically orthogonalized atomic
  orbital basis). The orthogonal basis functions are indexed by $r$,$s$,$t$,$u$
  (e.g., Eq.\,\eqref{eq:QCH}) and referred to as \emph{sites}. $Q$ is divided into
  sets of sites called \emph{fragments}, such that each site occurs in exactly
  one fragment. Fragments are handled sequentially, and for each fragment $A$,
  the union of the other fragments is treated as environment $B$. $|A|$ is
  the number of sites in set $A$.

  First, let us review the the exact formal bath that exactly embeds a given
  fragment $A$. Let $Q$ be the full quantum system, and $\ket{\Psi}$ be an
  eigenstate of $Q$.  $\ket{\Psi}$ may expanded as
  \begin{align} \ket{\Psi} = \sum\nolimits_{ij} \psi_{ij}
  \ket{\alpha_i} \ket{\beta_j} 
  \end{align} 
  where $\ket{\alpha_i}$ and $\ket{\beta_i}$ are states in the Fock space
  spanned by the fragment $A$ and the environment $B$, respectively. Simple
  algebra shows that
  \begin{align}
   \ket{\Psi} &= \sum\nolimits_{ij} \psi_{ij} \ket{\alpha_i} \ket{\beta_j} \notag
         = \sum\nolimits_{i} \ket{\alpha_i} \Big(\sum\nolimits_{j} \psi_{ij} \ket{\beta_j}\Big)\notag
  \\     &= \sum\nolimits_{i} \ket{\alpha_i} \ket{\tilde \chi_i}
         = \sum\nolimits_{ii'} \psi_{ii'} \ket{\alpha_i} \ket{\chi_{i'}}
  \label{eq:bath_decomposition}
  \end{align}
  where $\ket{\tilde \chi_i} =\sum\nolimits_{j} \psi_{ij} \ket{\beta_j}$, and
  $\ket{\chi_i}$ is the orthogonalized set of $\ket{\tilde \chi_i}$ states.
  (This rewriting is closely related to the Schmidt decomposition of quantum information
  theory\cite{peschel:EntangleMentInSolvableManyParticleModels}).
  Note that in the last line, although $\ket{\chi_{i'}}$ are states in the
  environment $B$, there are only $M_A$ of them: the dimension of the Fock space
  of $A$. This set of special environment states defines the bath.
  We see that
  (i) no matter how large the environment, in a given state $\ket{\Psi}$, a
  fragment $A$ can only be entangled with $M_A$ environment states in $B$. Thus
  the entanglement effect of the environment is \emph{fully represented by a
  bath of the same size as the fragment} it is embedding. This combination of
  the fragment with its bath, we refer to as the \emph{embedded} system.
  (ii) If $\ket{\Psi}$ is an eigenfunction of a Hamiltonian $H$ (that acts in
  the full system $Q$), it is also an eigenfunction of a projected Hamiltonian
  defined in only the embedded system, $H'=PHP$, where
  \begin{align}
   P&=\sum\nolimits_{ii'} \ket{\alpha_i} \ket{\chi_{i'}} \bra{\alpha_{i}} \bra{\chi_{i'}}. \label{eq:projector}
  \end{align}
  $H'$ is the \emph{embedding Hamiltonian}.

  We thus have established that fragments can, in principle, always be exactly
  embedded by baths no larger than the fragments themselves.  While exact, this
  result is purely formal, because to construct the bath states
  $\{\ket{\chi_{i}} \}$ we require knowledge of the state $\ket{\Psi}$ of the
  full system $Q$. However, it naturally suggests a practical approximation:
  construct the bath from an approximate state of $Q$, $\ket{\Phi}$, and use
  this approximate bath in a subsequent high-level treatment of the embedded
  fragments. This is the combination we aim for in DMET.

  A simple choice for $\ket{\Phi}$ is a Slater determinant (for example, as
  obtained from a mean-field treatment of the full system). For a Slater
  determinant, the associated bath and embedding Hamiltonian are particularly
  simple: they can be obtained from single-particle linear-algebra rather than
  the many-particle decomposition in Eq. \eqref{eq:bath_decomposition}, as we
  now show. First note that, for any $\ket{\Psi}$, the fragment many-body states
  $\ket{\alpha_i}$ in Eq. \eqref{eq:bath_decomposition} live in the Fock space
  spanned by the one-particle fragment sites $\mathcal{F}(\ket{i})$, $i\in A$.
  In the special case where $\ket{\Psi}=\ket{\Phi}$ is a determinant,
  also the bath states $\ket{\chi_i}$ live in a Fock space defined by
  one-particle bath orbitals $\{ \ket{b} \}$ (at most $|A|$), multiplied by a
  common core determinant. This is seen as follows. Let $\ket{p}=\sum_r C^r_p\ket{r}$
  denote the $N$ occupied orbitals of $\ket{\Phi}$, where $p=1 \ldots N$, $r \in Q$.
  Let
  \begin{align}
     S_{pq} = \sum\nolimits_{i \in A} \braket{p|i}\braket{i|q}
  \end{align}
  define the overlap matrix $S$ of the orbitals \emph{projected onto the sites
  of fragment $A$}. Then the eigenvectors of $S$ define a rotation of the
  occupied orbitals $\ket{p} \to \ket{\tilde{p}}$ which divides them into two
  sets: a set of $N-|A|$ occupied orbitals with zero eigenvalues,
  and thus \emph{without} any component on the fragment sites, and a set of
  $|A|$ occupied orbitals with non-zero eigenvalues, which have overlap
  with the fragment sites. We call the former ``pure environment orbitals'', and
  the latter ``entangled orbitals''. Projecting the entangled orbitals
  $\tilde{p}=1\ldots|A|$ onto the environment sites $B$ and normalizing then
  yields a set of bath orbitals $\ket{b}$ of the same number as fragment sites,
  \begin{align}
  \{\ket{b}\} = \bigg\{\frac{\sum\nolimits_{j \in B} \ket{j}\braket{j|\tilde{p}}}{\Vert\sum\nolimits_{j \in B} \ket{j}\braket{j|\tilde{p}}\Vert};\;\tilde{p}=1\ldots
  {A}\bigg\}.
  \end{align}
  Rewriting $\ket{\Phi}$ in terms of the rotated orbitals $\ket{\tilde{p}}$, and
  expressing each $\ket{\tilde{p}}$ in terms of its fragment, bath, and pure
  environment components, we see that the many-body bath states $\ket{\chi_i}$
  span the same space as
  $\mathcal{F}(\ket{b}) \otimes \mathrm{det}(e_1 e_2 \ldots e_{N-|A|})$,
  where
  $\mathrm{det}(e_1 e_2 \ldots e_{N-|A|})$ is the determinant of
  pure environment orbitals. In other words, when split across a fragment and
  environment, a determinant $\ket{\Phi}$ appears as a CAS-CI (complete active
  space configuration interaction) expansion in a half-filled active
  \emph{embedding} basis of fragment plus bath orbitals,
  $\{\ket{i}\} \oplus \{\ket{b}\}$,  with a core determinant of pure environment
  orbitals $\ket{e}$.

  \begin{figure*}
    \centering
    \includegraphics[width=.93\textwidth]{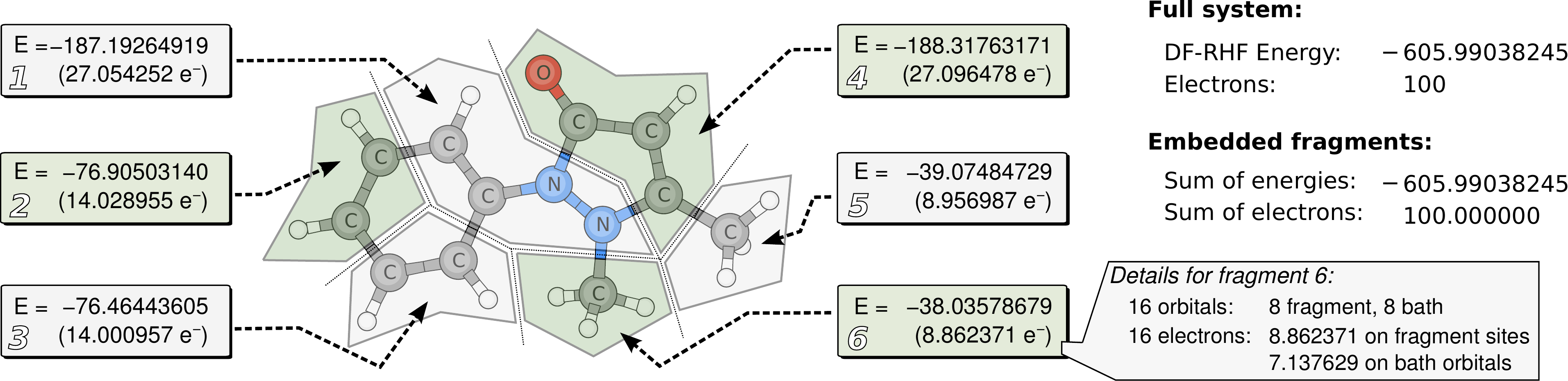}
    \caption{Example for HF in HF embedding: A molecule is split into random
      atomic fragments. For each fragment, an embedding is constructed, and the
      embedded system is treated with Hartree-Fock. The left panel shows the
      obtained distribution of energy and electrons amongst the fragments. The
      right inset shows how the open system embedding (e.g., a fractional
      electron number) is realized by coupling to a bath.}
    \label{fig:HFembedding}
  \end{figure*}

  We next construct the embedding Hamiltonian corresponding to the Slater
  determinant $\ket{\Phi}$. Formally, this is defined from the many-body
  projection Eq. \eqref{eq:projector}, but for the case of a Slater determinant,
  it can be constructed by  a simple change of single-particle basis. We define
  the embedded Hamiltonian $H'$ by projecting $H$ into the space
$\mathcal{F}(\ket{i}) \otimes \mathcal{F}(\ket{b}) \otimes \mathrm{det}(e_1 e_2 \ldots e_{N-|A|})$.
  This is equivalent to transforming $H$ into the active space of the embedding
  basis (fragment plus bath orbitals), and including a core contribution from
  the pure environment determinant, $\mathrm{det}(e_1 e_2 \ldots e_{N-|A|})$.
  Denoting the embedding basis by labels $v$, $w$, $x$, $y$, and its
  representation in terms of the full system sites by $\ket{x} = B^r_x\ket{r}$,
  we find
   \begin{align}\label{eq:EmbeddedH}
       H' &= \sum_{vw} h^v_w E^w_v + \frac{1}{2} \sum_{vwxy} V^{vw}_{xy} E^{xy}_{vw},
    \\ h^v_w &= B^r_v h^r_s B^s_w + f^\mathrm{core}_{vw}
    \\ V^{vw}_{xy} &= B^r_v B^w_s V^{rs}_{tu} B^t_x B^u_y.
   \end{align}
  If the full system $\ket{\Phi}$ was obtained from a Hartree-Fock calculation,
  then carrying out a Hartree-Fock calculation in this embedded system with the
  embedding Hamiltonian $H'$ yields a mean-field Fock operator $f'$ for which
  $\ket{\Phi}$ is an eigenstate.

  In \figref{fig:HFembedding} we numerically demonstrate the exactness of the
  above mean-field embedding by the following process: (i) a Hartree-Fock
  calculation is performed on a molecule, (ii) the molecule is split into
  arbitrary groups of atoms as fragments. For each fragment, an embedding is
  constructed (bath orbitals and $H'$) and a Hartree-Fock calculation is run on
  the embedded system, (iii) the system is reassembled by adding up the energies
  and electrons located on the various fragments. We note that only the
  electrons and energy contributions associated with the fragment sites should
  be considered when re-assembling the system, not the contributions associated
  exclusively with bath or pure environment orbitals, as this would lead to
  double counting. To that end, we define the energy of a fragment $A$ using the
  one- and two-particle density matrices, $\gamma$, $\Gamma$, with at least one
  index in fragment $A$,
  \begin{align}
     E_A &= \sum_{i \in A, s} \gamma^i_s h_i^s + \frac{1}{2}\sum_{i \in A, stu}\Gamma^{is}_{tu} V_{is}^{tu}.
  \end{align}
  Note that  $\gamma$ and $\Gamma$  include the contributions from the pure
  environment orbitals (the core determinant). As we see from
  \figref{fig:HFembedding}, the embedding allows an arbitrary fragmentation into
  open fragments to be exactly reassembled (note the fractional electron
  numbers!), recovering the exact Hartree-Fock energy of the wave function
  $\ket{\Phi}$ used to contruct it.

  We now return to the DMET. Recall that here we still construct the embedding
  based on a mean-field $\ket{\Phi}$ of the full system $Q$, but use it to
  embed high level, calculations on the fragments, rather than Hartree-Fock. For
  each fragment $A$, the high level calculation on the embedded system (fragment
  sites, bath orbitals, and Hamiltonian $H'$) yields a correlated
  $\ket{\Psi_A}$. Here we require an additional self-consistency cycle to ensure
  a consistent fragment description by the mean-field $\ket{\Phi}$ and the
  high-level embedded $\ket{\Psi_A}$. Although the choice of self-consistency
  condition is not unique, we note that a mean-field state is characterized by
  its one-particle density matrix $\braket{\Phi|a^\dag_r a_s|\Phi}$, and thus it
  is convenient to enforce consistency at the level of the density matrices by
  minimizing the metric
  \begin{align}\label{eq:FittingMetric}
  \Delta =  \sum_A \sum_{rs \in A} \Vert\braket{\Phi|a^\dag_r a_s|\Phi} - \braket{\Psi_A|a^\dag_r a_s|\Psi_A}\Vert^2.
  \end{align}
  Here we define the density matrix difference only over (intra)fragment sites
  for simplicity, but very similar results are obtained by defining the
  difference over fragment plus bath sites for each embedding, as we did in our
  earlier work.\cite{knizia:DmetHubbard} $\ket{\Phi}$ is then obtained from a
  mean-field Fock operator $f$ augmented by a set of one-particle operators
  $u_A$ for each fragment, with $u^A_{rs}=0$ for $r\notin A$ or $s\notin A$. The
  $u_A$ are chosen to minimize $\Delta$ and capture the correlation effects on
  the one-particle density matrix. The embedded Hamiltonian is also augmented
  with the correlation operators on the fragments other than the one currently
  being considered, projected into the embedding basis.
  The DMET self-consistency cycle is thus:
  \begin{enumerate}
   \item The full system is treated at the Hartree-Fock level using the Fock
      operator $f + \sum_A u_A$ to determine $\ket{\Phi}$. Initially all ${u}_A$
      are zero.
   \item For each fragment, $\ket{\Phi}$ is used to construct an embedding
      basis. The embedding Hamiltonian for fragment $A$, $H_A'$, is obtained by
      projecting $H+\sum_{A'\neq A} u_{A'}$ into the embedding basis following
      Eq.\,\eqref{eq:EmbeddedH}. The embedded fragment's state $\ket{\Psi_A}$ is
      calculated at a correlated level, for example, with full configuration
      interaction (FCI).
    \item For each fragment $A$, we adjust the correlation operator
          $u_A = \sum_{rs\in A} u^A_{rs} E_r^s$
          to minimize the difference between the Hartree-Fock one-particle
          density matrix and the correlated one-particle density matrix,
          $\Delta_A=\sum_{rs\in A}\Vert\braket{\Phi|a^\dag_r a_s|\Phi} - \braket{\Psi_A|a^\dag_r a_s|\Psi_A}\Vert^2$.
    \item The cycle is iterated until all $u^A_{rs}$ converge.
  \end{enumerate}

  \begin{figure}
      \centering
      \includegraphics[width=.95\columnwidth]{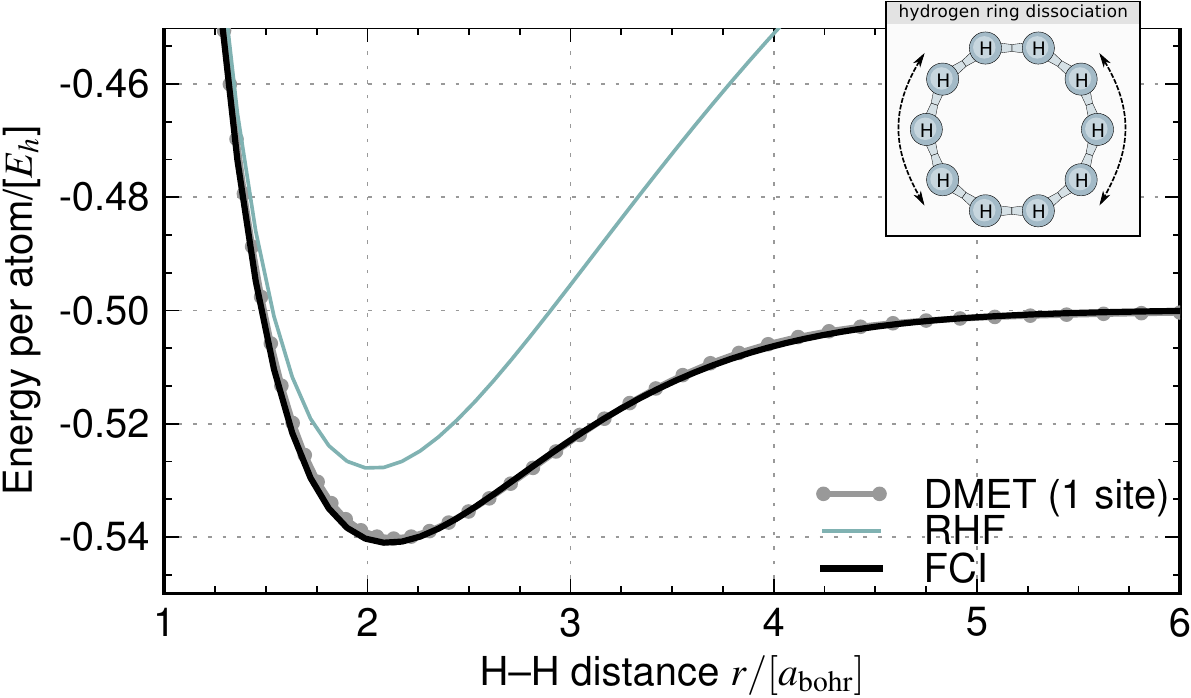}
      \caption{Symmetric stretching of a H$_{10}$ ring: Shown are the exact
         results (FCI), the mean-field results (RHF), and the 1-site DMET
         results. In 1-site DMET, the system is fragmented into individual atoms
         treated with FCI, and embeddings are constructed from full system RHF.}
      \label{fig:H10ring}
  \end{figure}

  In order to test whether the DMET provides reasonable results, we now consider
  simple model systems exhibiting strong correlation; namely, hydrogen rings,
  chains, and grids. Such systems have recently emerged as a rich testbed for
  new correlation methods, as the strength of the correlation can be readily
  tuned from weak to strong by changing the hydrogen atom spacing.%
  \cite{tsuchimochi:cpmft,Mitrushchenkov:DmrgOrbitalOpt,Stella:AfqmHydrogenChains,alsaidi:BondBreakingWithAfqmc,bytautas:SeniorityHierarchies,Lin:DmftForQC}
  Note that such systems show high degeneracy and are difficult to handle even
  with full correlation treatments. Furthermore, as we argued in the
  introduction, any choice of fragment will be strongly coupled to the rest of
  the system, so potential based embeddings cannot describe them. All
  calculations employ a minimal hydrogen basis consisting of orthogonalized
  1$s$-like AO functions obtained from an underlying cc-pVTZ basis, except for
  the H$_{50}$ calculation, for which we use a STO-6G basis to retain
  compatibility with earlier references. For the high level treatment of the
  embedded systems, we use FCI. All calculations are spin-adapted, and both the
  mean-field (restricted HF) and correlated (FCI) calculations use singlet wave
  functions. As DMET self consistency metric we used
  Eq.\;\eqref{eq:FittingMetric}, except for the 3-site calculation on the
  $4\times3$ grid, where the previous metric\cite{knizia:DmetHubbard} was used
  to avoid convergence problems at $r>3.8 a_\mathrm{bohr}$.

  As the simplest non-trivial example, we investigate the symmetric dissociation
  of a ring of ten hydrogen atoms (that is, all bonds are simultaneously
  stretched). The system is fragmented into individual 1$s$ orbitals of the
  hydrogen atoms, and thus each site is one fragment. The results are shown in
  \figref{fig:H10ring}. The 1-site DMET calculation almost exactly reproduces
  the reference FCI curve. Note that each correlated fragment calculation
  corresponds to a FCI calculation on only two orbitals: the H 1$s$ orbital, and
  a single bath orbital, and is thus a trivial $3\times 3$ matrix
  diagonalization.%
  \footnote{Two orbitals span four two-electron states, of which three are singlet.}

  As second example, we choose an inhomogeneous system: the linear H$_{50}$
  chain. FCI on this system would require on the order of $10^{28}$
  determinants, so truly exact results cannot be calculated. However, near-exact
  reference data can be obtained by the quantum chemistry density matrix
  renormalization group (DMRG).\cite{chan:DmrgInQcReview} We here take the data
  from Hachmann et al.\cite{hachmann:DmrgForLongMolecules} This particular
  system has also been the subject of a recent DMFT study by Lin et
  al.,\cite{Lin:DmftForQC} the results of which are shown for comparison. It is
  clear that the 1-site DMET once again closely reproduces the reference values.
  Again the fragment calculations are two orbital FCI and thus numerically
  trivial. It is noteworthy that in this case DMET does better than the more
  complex DMFT, likely due to its ability to treat long-range interactions with
  the bath beyond mean-field.

  \begin{figure}
      \centering
      \includegraphics[width=.95\columnwidth]{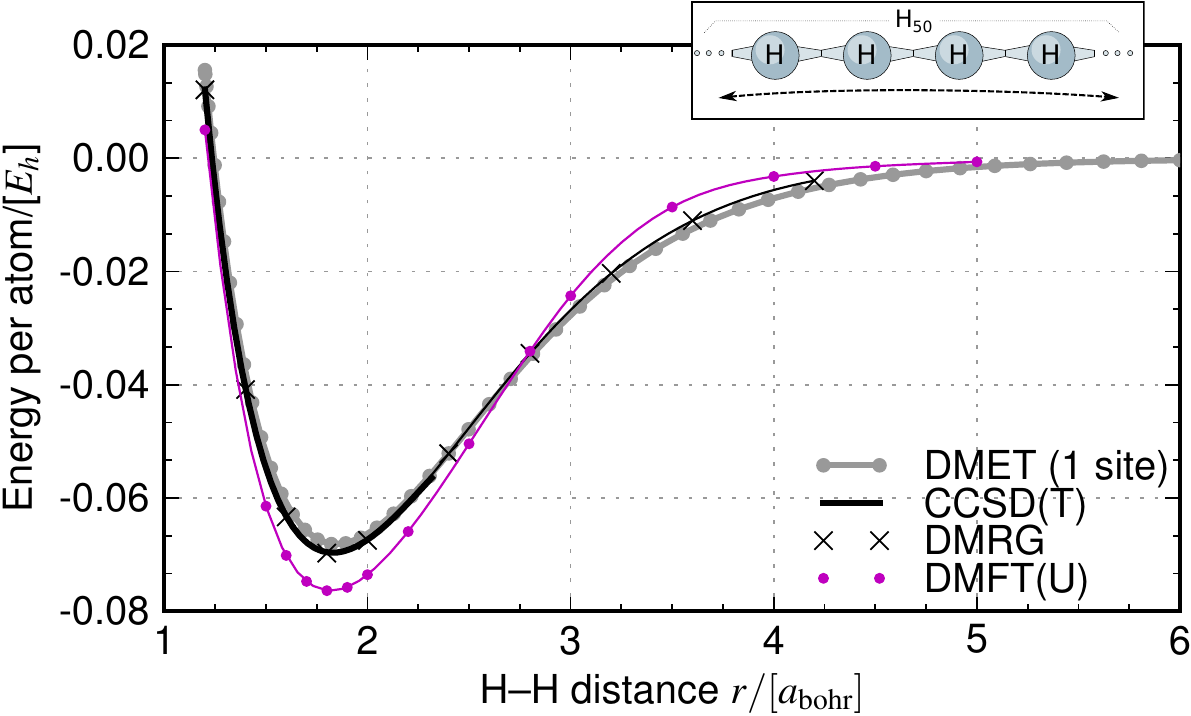}
      \caption{Symmetric stretching of a H$_{50}$ chain: Shown are reference
         DMRG\cite{hachmann:DmrgForLongMolecules} and CCSD(T) data, the
         mean-field results (RHF), the 1-site DMET results, and the DMFT results
         by Lin et al.\cite{Lin:DmftForQC}}
      \label{fig:H50chain}
  \end{figure}

  \begin{figure}
      \centering
      \includegraphics[width=.95\columnwidth]{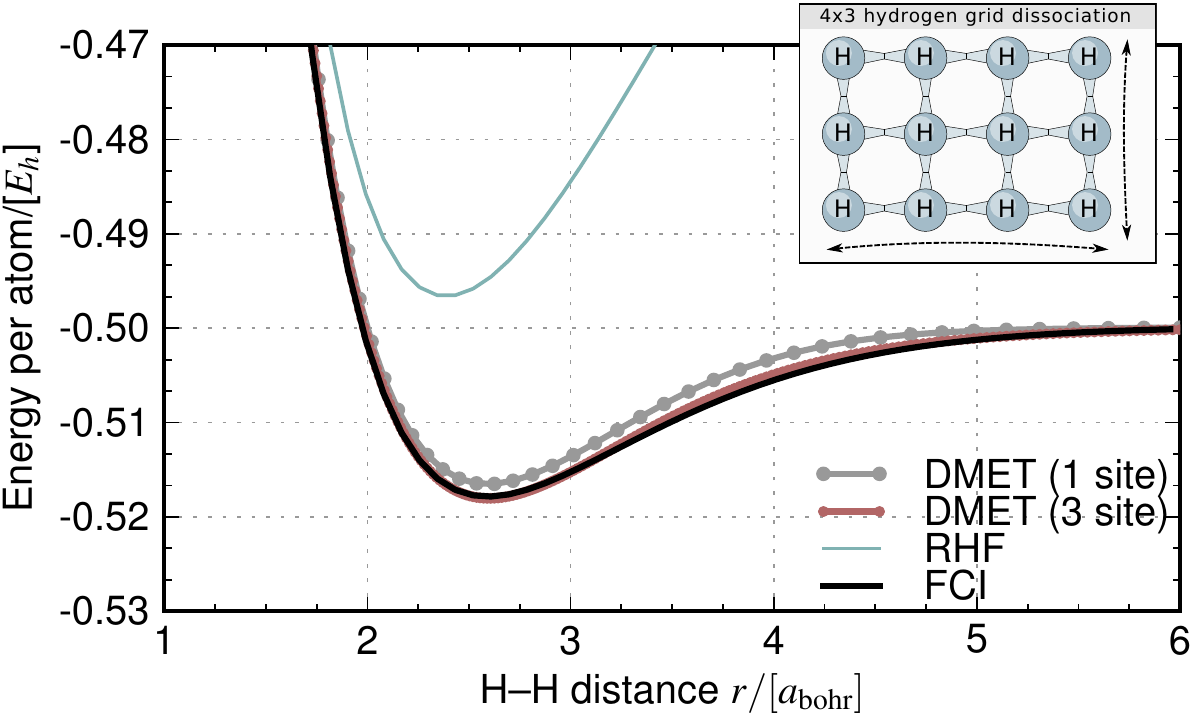}
      \caption{Symmetric stretching of a $4\times 3$ hydrogen grid. In the
         3-site DMET, the system is fragmented into four columns of three
         hydrogens each.}
      \label{fig:H12grid}
  \end{figure}

  For a more challenging system, we now turn to a two-dimensional inhomogeneous
  system, the $4\times 3$ hydrogen grid. This system is very pathological---it
  is non-binding at Hartree-Fock level, and converging the normally \emph{very}
  robust FCI\cite{knowles:fci} required hundreds of iterations for some points.
  As shown in \figref{fig:H12grid}, even here the 1-site DMET qualitatively
  reproduces the FCI binding curve. The calculations remain as trivial to
  perform as for the one-dimensional systems. Additionally, if we embed entire
  columns of atoms, carrying out a 3-site DMET, the agreement between the
  embedded and the reference results becomes almost perfect.

  The accuracy of these results may seem surprising, given that the systems are
  strongly correlated but the embedding is obtained from an uncorrelated,
  qualitatively incorrect, mean-field $\ket{\Phi}$. However, it is only the bath
  \emph{states} that are determined from the mean-field theory. Once those
  states are determined, the embedded Hamiltonian is constructed exactly, and
  the coupling between the fragment sites and the bath orbitals is obtained by
  FCI, not mean-field. This allows the fragment state to correctly transition
  from the delocalized regime of weak correlation to the entangled spin regime
  of strong correlation. As long as the entanglement is reasonably local and
  does not involve the \emph{pure} environment orbitals,  we can expect good
  results.

  The robustness and simplicity of DMET, even in the presence of strong coupling
  and strong correlation, makes it unique amongst current embedding approaches,
  and suggest that it could be useful in a wide range of applications. The next
  step will be to apply the theory to  more realistic and larger scale chemical
  problems. We are now pursuing these studies.

  This work was supported by the Department of Energy, Office of Science,
  through Grant No. DE--FG02--07ER46432 and the Computational Materials Science
  Network (DE-SC0006613).


%

\end{document}